# Synthesis of Large-Area MoS$_2$ Atomic Layers with Chemical Vapor Deposition

Yi-Hsien Lee#, Xin-Quan Zhang#, Wenjing Zhang, Mu-Tung Chang, Cheng-Te Lin, Kai-Di Chang,Ya-Chu Yu, Jacob Tse-Wei Wang, Chia-Seng Chang, Lain-Jong Li* and Tsung-Wu Lin*

Transition metal dichalcogenides (TMD), MX$_2$ (M=Mo, W; X=S, Se, Te), have attracted considerable attention for their great potential in the fields of catalysis, nanotribology, microelectronics, lithium batteries, hydrogen storage, medical and optoelectronics.[1-12] MoS$_2$ nano-materials have been known in the form of nested fullerene-like nanodots and one-dimensional nanotubes.[1-4,13-17] Stimulated by the discovery of two-dimensional graphene monolayer and its rich physical phenomenon, inorganic graphene analogues such as layered MoS$_2$, where the Mo layer is sandwiched between two sulfur layers by covalent forces, have created great interest in the past few years. Recently, Radisavljevic *et al.* have demonstrated that the transistors fabricated with the exfoliated MoS$_2$ monolayer[18-19] exhibit high on-off current ratio and good electrical performance, which may be used in future electronic circuits requiring low stand-by power. The strong emission inherited from the direct gap structure of monolayer MoS$_2$ also promises the applications in optoelectronics.[20-22]

Substantial efforts have been devoted to prepare thin-layer MoS$_2$, including scotch tape based micromechanical exfoliation,[18-24] intercalation assisted exfoliation,[25-27] liquid exfoliation,[28] physical vapor deposition,[29-230] hydrothermal synthesis,[31] thermolysis of single precursor containing Mo and S.[32-33] The lateral size of the MoS$_2$ films synthesized by the aforementioned methods is often in the order of several micrometer; however, the synthesis of large-size MoS$_2$ thin layers is still a challenge. Chemical vapor deposition (CVD) has been one of the most practical methods for synthesizing large-area graphene[34-36] and graphene analogues such as boron nitride and BCN nanosheets.[37-38] The sulfurization of MoO$_3$ using the CVD method has been adopted to synthesize MoS$_2$ materials; however, the reaction normally leads to MoS$_2$ nanoparticles or nanorod structures during the synthesis.[39-40] To our best knowledge, synthesis of large-area, monolayer MoS$_2$ films on amorphous SiO$_2$ substrates using a CVD method has not yet been reported. In this contribution, CVD is adopted to synthesize MoS$_2$ layer directly on SiO$_2$/Si substrates using MoO$_3$ and S powders as the reactants. The growth of MoS$_2$ is very sensitive to the substrate treatment prior to the growth. The use of graphene-like molecules for the substrate treatment, such as reduced graphene oxide (rGO), perylene-3,4,9,10-tetracarboxylic acid tetrapotassium salt (PTAS) and perylene-3,4,9,10-tetracarboxylic dianhydride (PTCDA), promotes the layer growth of MoS$_2$. Large-area MoS$_2$ layers can be directly obtained on amorphous SiO$_2$ surfaces without the need to use highly crystalline metal substrates or an ultrahigh vacuum environment, which is in clear contrast to the reported epitaxial growth of MoS$_2$ nano-islands on crystalline Au(111) surfaces in ultrahigh vacuum.[29] Spectroscopic, microscopic and electrical measurements suggest that the synthetic process leads to the growth of monolayer, bilayer and few-layer MoS$_2$ sheets. These MoS$_2$ films are highly crystalline and their size is up to several millimeters.

Figure 1a schematically illustrates our experimental set-up. The MoO$_3$ powder (0.4 g) was placed in a ceramic boat and the SiO$_2$/Si substrate was faced down and mounted on the top of boat. A separate ceramic boat with sulfur powder (0.8 g) was placed next to the MoO$_3$ powder. Prior to the growth, a droplet of aqueous reduced graphene oxide (rGO), PTAS or PTCDA solution, was spun on the substrate surface followed by drying at 50 $^o$C. During the synthesis of MoS$_2$ sheets, the reaction chamber was heated to 650$^o$C in a nitrogen environment. At such a high temperature, MoO$_3$ powder was reduced by the sulfur vapor to form volatile suboxide MoO$_{3-x}$.[39] These suboxide compounds diffused to the substrate and further reacted with sulfur vapor to grow MoS$_2$ films. Figure 1b displays the OM of the MoS$_2$ sheets obtained on the SiO$_2$/Si substrate pretreated with an rGO solution and inset shows that a white dot is present at the center of a star-shaped MoS$_2$ sheet, where these dots seem to act as the seeds for growing MoS$_2$ layers. More images are shown in supporting figure S1 to evidence the observation. The star-shaped MoS$_2$ can be merged to form a continuous MoS$_2$ film (with a lateral size up to 2 mm) as shown in the upper area of figure 1b, where the seed density is higher. In figure 1c, smooth surface morphology of MoS$_2$ sheets is observed with atomic force microscope (AFM), suggesting that a layer structure of MoS$_2$ is formed. The cross-sectional height in figure 1d

[*] Dr. Y.-H. Lee, Dr. W. Zhang, Dr. C.-T. Lin, Mr. J. T.-W. Wang, Dr. L.-J. Li*
*Institute of Atomic and Molecular Sciences, Academia Sinica,*
*128 Sec. 2, Academia Rd., Nankang, Taipei 11529, Taiwan*
Fax:(+886) 35734217
E-mail: lanceli@gate.sinica.edu.tw

Dr. M.-T. Chang, Dr. C. S. Chang
*Institute of Physics, Academia Sinica,*
*128 Sec. 2, Academia Rd., Nankang, Taipei 11529, Taiwan*
Mr. X.-Q. Zhang, Mr. K.-D. Chang, Ms. Y.-C. Yu, Dr. T.-W. Lin*
*Department of Chemistry, Tunghai University,*
*No. 181, Sec. 3, Taichung Port Rd., Taichung City 40704, Taiwan*
Fax: (+886) 423596233
E-mail: twlin@thu.edu.tw

[**] # These authors contributed equally. This research was supported by National Science Council Taiwan (NSC 100-2113-M-029-001-MY2, NSC-99-2112-M-001-021-MY3 and NSC 100-2113-M-029-001-MY2), RCAS and Academia Sinica. We also acknowledge the support from National Nano Projects (NSC), NCTU and NTHU in Taiwan.

reveals that the thickness of the MoS$_2$ film is ~0.72 nm, which corresponds to a monolayer MoS$_2$ sheet based on previous reports for a monolayer MoS$_2$ on a Si/SiO$_2$ substrate.[26] In addition to monolayer MoS$_2$, we also occasionally find bilayer, trilayer and other thicker layers. Supporting figure S2 provides AFM images and cross-sectional profiles of the thicker MoS$_2$ films. Supporting figure S3 shows the optical micrographs (OM) of the layered MoS$_2$ grown respectively with the PTAS and PTCDA pre-treatments, where the MoS$_2$ layer growth is initiated from the PTAS or PTCDA molecular aggregates. Similar to the role of rGO, the PTAS or PTCDA molecular aggregates act as the seeds for growing MoS$_2$ thin layers. Experimentally, the MoS$_2$ layer growth initiated by rGO is more homogenous in layer thickness. Hence, the subsequent discussions are mainly based on rGO initiated MoS$_2$ thin layers.

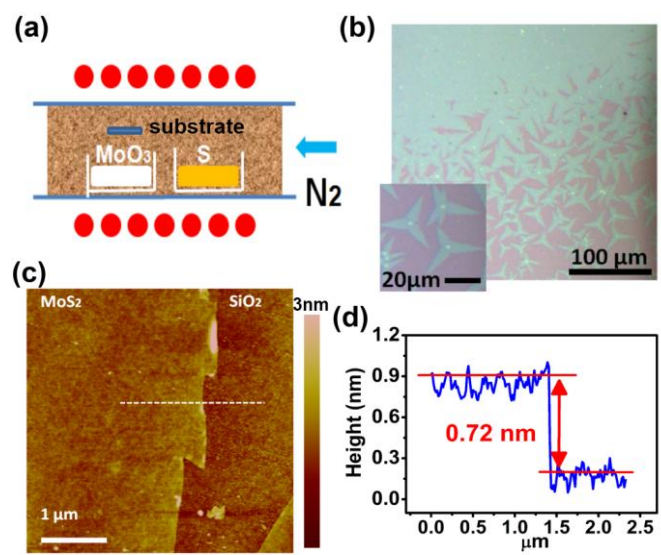

**Figure 1.** (a) Schematic illustration for the experimental set-up. (b) The optical micrographs of the MoS$_2$ layers grown on the substrate respectively treated with rGO solution. The inset shows the magnified OM of the MoS$_2$ films, where the seed is observed at the center of each star-shaped sheet. (c) AFM image of a monolayer MoS$_2$ film on a SiO$_2$/ Si substrate (pre-treated with rGO). (d) The thickness of the MoS$_2$ layer is 0.72 nm from the AFM cross-sectional profile along the line indicated in (c).

To explore the Raman and PL dependency on MoS$_2$ layer thickness, we identify an area with MoS$_2$ monolayer, bilayer and trilayer films. Figure 2a and 2b respectively shows the mapping constructed by plotting the integrated MoS$_2$ Raman peak intensity (360 ~ 420 cm$^{-1}$) and the PL peak intensity (650 ~ 700 nm) in confocal measurements. The thickness distribution seems to correlate well to the contrast in OM image figure 2c. The MoS$_2$ monolayer sheet exhibits two Raman characteristic bands at 403.8 and 385.8 cm$^{-1}$ with the full-width-half-maximum (FWHM) values of 6.6 and 3.5 cm$^{-1}$, corresponding to the A$_{1g}$ and E$_{2g}$ modes respectively. Note that the peak frequency difference between A$_{1g}$ and E$_{2g}$ modes ($\Delta$) can be used to identify the layer number of MoS$_2$. The value of $\Delta$ (18 cm$^{-1}$) in figure 2d evidences the existence of monolayer MoS$_2$.[24,41] The inset in Figure 2d shows that the $\Delta$ value increases with the layer number of MoS$_2$, where the layer number is confirmed by AFM thickness (Supporting figure S2). These results agree well with the observation for exfoliated MoS$_2$ layer.[23] In figure 2e, the PL spectrum shows two pronounced emission peaks at 627 and 677 nm[24] and these emissions are known as the A1 and B1 direct excitonic transitions.[44] The emission intensity (normalized by the Raman scattering at ~482 nm) obviously decreases with the layer number. This can be reasoned by the fact that the optical bandgap transforms from indirect to direct one when the dimension of MoS$_2$ is reduced from a bulk form to a monolayer sheet.[19] The X-ray photoelectron spectroscopy (XPS) scans for the monolayer MoS$_2$ sample confirm the chemical bonding states of the MoS$_2$ layers (Supporting figure S4). These binding energies are all consistent with the reported values for MoS$_2$ crystal.[32,43]

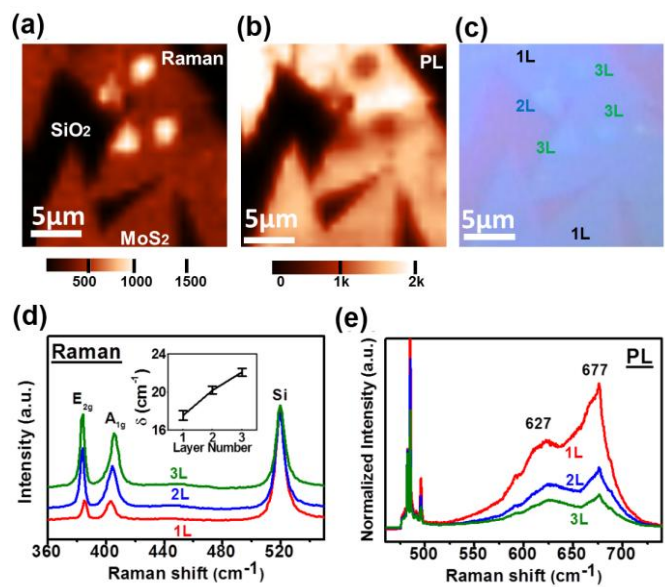

***Figure 2*** (a) Raman peak intensity mapping (360 ~ 420 cm$^{-1}$), (b) PL peak intensity mapping (650 ~ 700 nm) and (c) OM image of the selected area with various MoS$_2$ layer thickness (1L, 2L and 3L). (d) Raman spectra and (e) photoluminescence of the monolayer, bilayer and trilayer MoS$_2$ sheets. Both Raman and PL experiments were performed in a confocal spectrometer using a 473 nm excitation laser.

Figure 3a shows the transmission electron microscopy (TEM) image for the monolayer MoS$_2$. The high resolution TEM image (figure 3b) and the corresponding selected area electron diffraction (SAED) pattern with [001] zone axis (inset of figure 3b) reveal the hexagonal lattice structure with the lattice spacing of 0.27 and 0.16 nm assigned to the (100) and (110) planes. The distinct SAED pattern suggests that the crystalline domain of the MoS$_2$ layer is at least 160 nm in lateral size (SAED aperture size ~160 nm in our measurement). Figure 3c displays the TEM image for the selected grain boundary area as indicated by the inset AFM, where the junction between two MoS$_2$ domains is clearly seen. The in-plane X-ray diffraction (XRD) profile for the MoS$_2$ monolayer synthesized by the CVD method is shown in Figure 3d and the diffraction peaks at 32.4 and 58 degree are attributed to the (100) and (110) crystal planes respectively. Meanwhile, the stoichiometry of the MoS$_2$ film has been separately confirmed with XPS (S/Mo ratio ~ 2.xx) and energy dispersive TEM based X-ray spectroscopy (EDS) as shown in supporting figure S5.

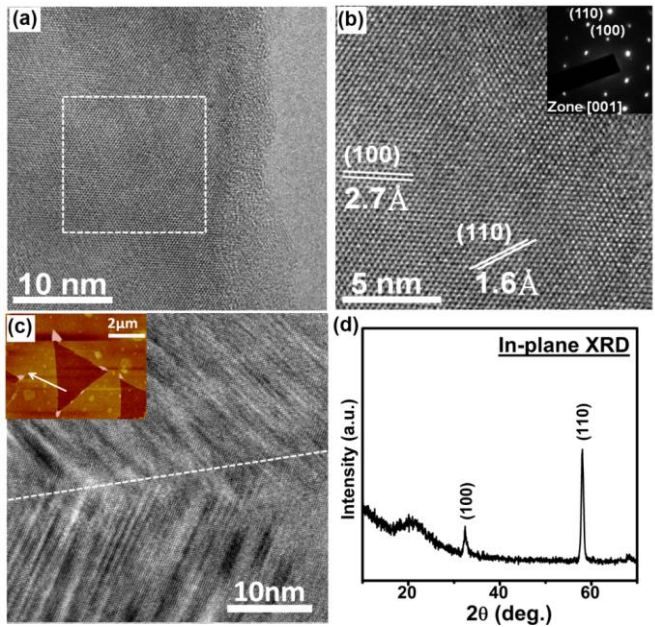

**Figure 3.** (a) High resolution TEM image of $MoS_2$ monolayer. (b) Enlarged HR-TEM image of the marked area in figure (a) with an inset showing the SAED pattern. (c) TEM image for the $MoS_2$ domain boundary at the location as indicated by the inset AFM image. (d) In-plane XRD result for the $MoS_2$ monolayer.

To evaluate the electrical performance of the $MoS_2$ sheets, we fabricate bottom-gated transistors on $SiO_2$/Si using conventional photolithography. The bottom-gate transistors were fabricated by evaporating Au electrodes directly on top of the $MoS_2$ layer. Figure 4 demonstrates the transfer curve (drain current $I_d$ vs. gate voltage $V_g$) for the device prepared from a $MoS_2$ monolayer. Inset shows the top view OM of the device. The on-off current ratio is approximately $1 \times 10^4$. The field-effect mobility of holes was extracted based on the slope $\Delta I_d/\Delta V_g$ fitted to the linear regime of the transfer curves using the equation $\mu = (L/WC_{ox}V_d)(\Delta I_d/\Delta V_g)$, where L, W and $C_{ox}$ are the channel length, width and the gate capacitance.[44] The effective field effect mobility for the $MoS_2$ device can be up to 0.02 $cm^2$/(V-s) in ambient, in agreement with previous reports.[18,41,45-46] We note that the valley point of the transfer curve is at -84V and the FET shows the typical n-type behavior, which is consistent with other reports.[18,47] Although the device exhibits a reasonably high on/off current ratio, there is still room to improve the carrier mobility. The relatively lower carrier mobility than the mechanically exfoliated $MoS_2$ is likely limited by the structural defects, such as the grain boundary observed by TEM (Figure 3c).

As revealed in figure 2b, the star-shaped $MoS_2$ layers were grown from center seeds, which suggest that the nucleation was a crucial step. The spin-casting of rGO solution before CVD growth introduced some tiny rGO flakes on the substrate surfaces, which experimentally enhanced the growth of $MoS_2$ layers. Supporting figure S6 and Table S1 shows that the morphology of the synthesized $MoS_2$ film is significantly affected by surface treatments. Without treating the substrate surface with rGO solution, only $MoS_2$ particles were found on the substrate. Other control experiments where the substrates separately cast with a graphene oxide (GO), hydrazine or KCl solution show that no $MoS_2$ layers but only sparsely distributed $MoS_2$ nano-particles are observed on substrates. Compared with more ordered aromatic structures of the graphene-like molecules including rGO, PTAS, and PTCDA, the GO is with randomly distributed defects and dangling bonds, which might be one of the reasons not being able to initiate layer growth. Although the GO may be thermally reduced to rGO[ref:pls add the J. Am. Chem. Soc. 2011, 133, 18522] at the $MoS_2$ growth temprature (650°C), the formation of $MoS_2$ seeds should involve many other factors such as the reaction between $MoO_3$ and S, the attachment of $MoO_{3-x}$ vapors onto GO, the conversion of $MoO_{3-x}$ to $MoS_2$, and the morphology of the $MoS_2$ seeds formed on substrates. These reactions may occur during the temperature ramping period. It is likely that the $MoS_2$ seed morphology formed on GO prefers particle growth rather than layer growth. It is noted that our experimental results only allow us to conclude that the rGO treatment helps to form the $MoS_2$ seeds which prefers and promotes the layer growth of $MoS_2$. The morphology and structure of the seeds, requiring intense research efforts, are currently under investigation in our group. Meanwhile, we observe that both $MoS_2$ and $WS_2$, two typical transition metal dichalcogenides (TMD), exhibits similar layer growth behavior on the substrates pre-treated with graphene-like molecules (Supporting figure S7). The growth of $MoS_2$ and $WS_2$ layers is highly reproducible with our experimental conditions. A similar enhancement is expected to be observed in other transition-metal-disulfide TMD family materials.

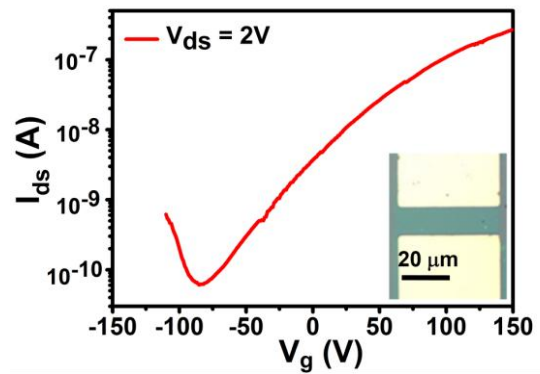

**Figure 4.** The typical transfer $I_d$-$V_g$ curve for a monolayer $MoS_2$ synthesized at 650 °C. The inset shows the OM imge of a FET device.

In conclusion, large-area $MoS_2$ films are directly synthesized on $SiO_2$/Si substrates with chemical vapor deposition. It is noteworthy that the growth of $MoS_2$ is not unique to $SiO_2$ substrates and it is also observed on other insulating substrates such as sapphire. The as-synthesized films are consisted of monolayer, bilayer and other few-layer $MoS_2$. Chemical configurations, including stoichiometry and valence states of $MoS_2$ layers are confirmed with XPS. Raman spectra and PL performance of the monolayer $MoS_2$ are presented. TEM and SAED demonstrate that the monolayer $MoS_2$ exhibits six-fold symmetry hexagonal lattice and high crystallinity. The electric measurement for the bottom-gate transistor shows a N-type semiconductor behaviour and the on-off current ratio is approximately $1 \times 10^4$. The seeding approach can be further used to grow other transition metal dichalcogenides.

*Experimental Section*

Synthesis: The $MoS_2$ films were synthesized on $SiO_2$/Si substrates in a hot-wall furnace. The ultra large and single-layer GO nanosheets are prepared by a modified Hummers' method as reported.[48] For the reduction of GO, GO solution was mixed with hydrazine solution and mixed solution was then heated to 90 °C for 1 hr.[49] Prior to the growth, a drop of rGO-hydrazine solution was spun on the substrate. PTAS (50 μM) or PTCDA (26mg in 5mL DI water) solution can also be used to treat the substrate. High purity $MoO_3$ (99%, Aldrich) and S powder (99.5, Alfa) were placed in two separate $Al_2O_3$ crucibles and the substrates were faced down and placed on the upper side of $MoO_3$ power. The $MoS_2$ samples were fabricated by annealing at 650 °C for 15 minutes with a heating rate of 15 °C /min and $N_2$ flow (1 sccm) at ambient.

Characterizations: Surface morphology of the samples was examined with commercial atomic force microscope (AFM, Veeco Icon) and scanning electron microscope (SEM, FEI VS600). Raman spectra and photoluminescence (PL) were obtained by confocal Raman microscopic systems (NT-MDT). Wavelength and spot size of the laser are 473 nm and 0.4 μm, respectively. The Si peak at 520 cm$^{-1}$ was used for calibration in the experiments. Field-emission transmission electron microscope (JEOL JEM-2100F, operated at 200 kV with a point-to-point resolution of 0.19 nm) equipped with an energy dispersive spectrometer (EDS) was used to obtain the information of the microstructures and the chemical compositions. The TEM samples were prepared using lacy-carbon Cu grid to scratch the surface of $MoS_2$ sample. Due to that only van der Waals force exists between $MoS_2$ and underlying SiO2 substrate, a few $MoS_2$ flakes may easily attach to the lacy-carbon TEM grid. Chemical configurations were determined by X-ray photoelectron spectroscope (XPS, Phi V5000). XPS measurements were performed with an Al Kα X-ray source on the samples. The energy calibrations were made against the C 1s peak to eliminate the charging of the sample during analysis.

*Supporting Materials*

# Synthesis of Large-Area MoS$_2$ Atomic Layers with Chemical Vapor Deposition

*Yi-Hsien Lee, Xin-Quan Zhang, Wenjing Zhang, Mu-Tung Chang, Cheng-Te Lin, Kai-Di Chang, Ya-Chu Yu, Jacob Tse-Wei Wang, Chia-Seng Chang, Lain-Jong Li\* and Tsung-Wu Lin\**

*Institute of Atomic and Molecular Sciences, Academia Sinica, 128 Sec. 2, Academia Rd., Nankang, Taipei 11529, Taiwan*

*Institute of Physics, Academia Sinica, 128 Sec. 2, Academia Rd., Nankang, Taipei 11529, Taiwan*

*Department of Chemistry, Tunghai University, No. 181, Sec. 3, Taichung Port Rd., Taichung City 40704, Taiwan*

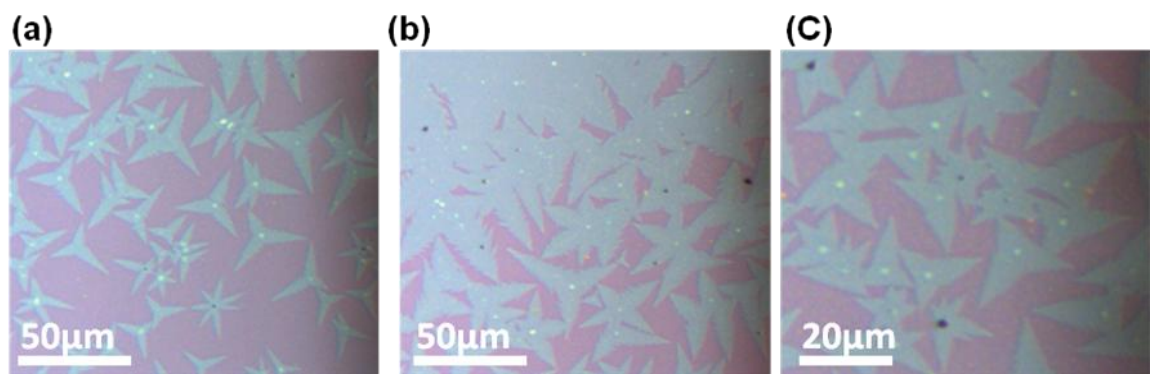

**Figure S1.** Optical micrographs showing the presence of center seeds of MoS$_2$ growth on the substrate pre-treatment with an rGO solution.

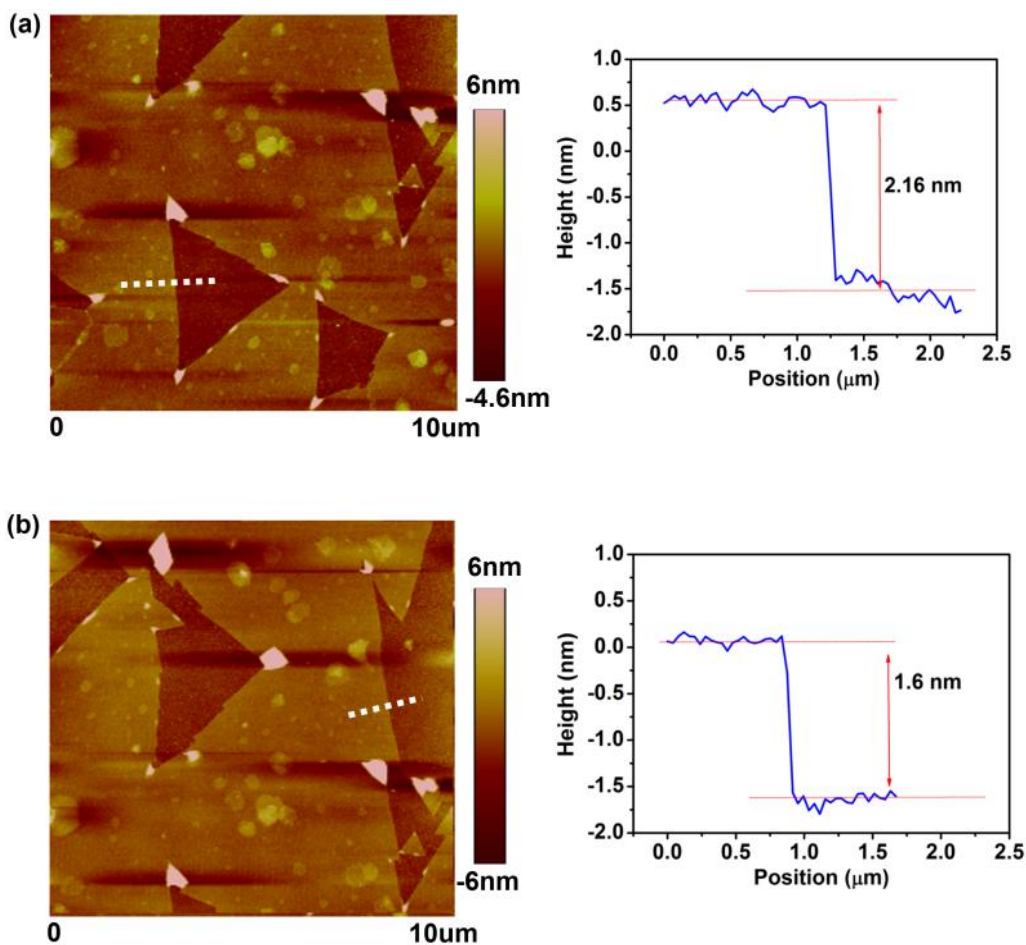

**Figure S2.** AFM images and cross-sectional profiles of the MoS$_2$ films with different thicknesses. (a) bilayer (b) trilayer MoS$_2$ film.

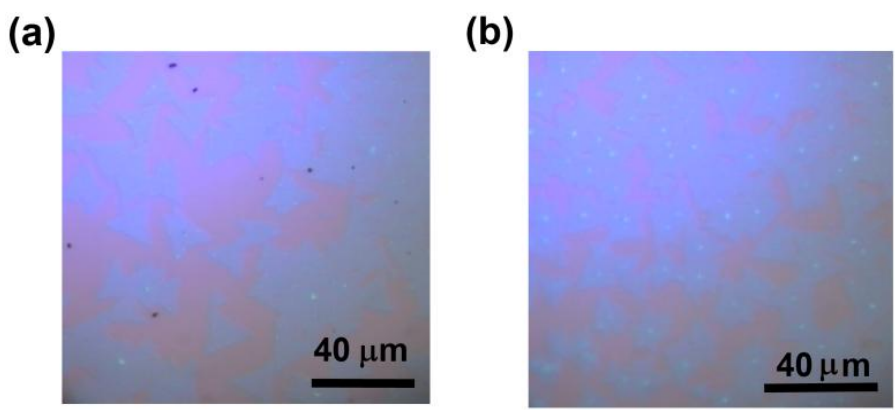

**Figure S3**. The optical micrographs (OM) of the layered MoS$_2$ grown respectively with the (a) PTAS and (b) PTCDA pre-treatments, where the MoS$_2$ layer growth is initiated from the PTAS or PTCDA molecular aggregates.

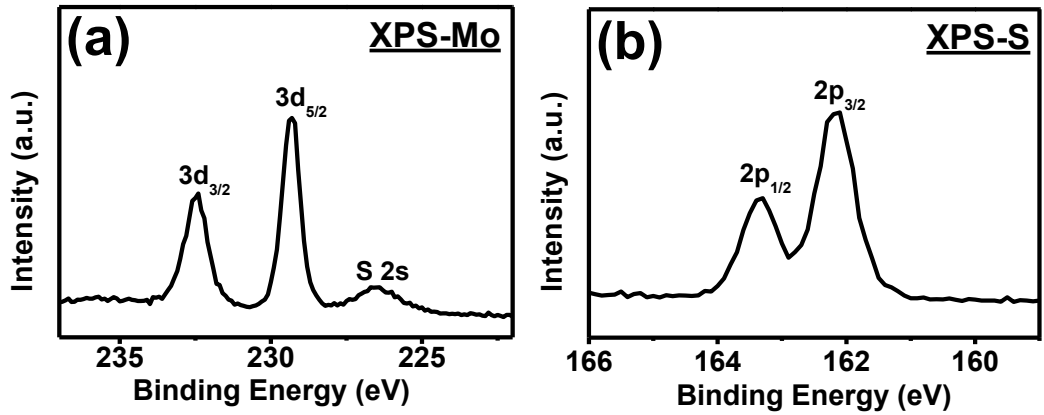

**Figure S4**. XPS spectra of the MoS2 film for (a) Mo 3d, where the two peaks at 229.3 and 232.5 eV, attributed to the doublet $Mo3d_{5/2}$ and $Mo3d_{3/2}$. (b) S 2p, where the binding energy at 162.2 and 163.3 eV can be assigned to spin-orbit $S2p_{3/2}$ and $S2p_{1/2}$, respectively.

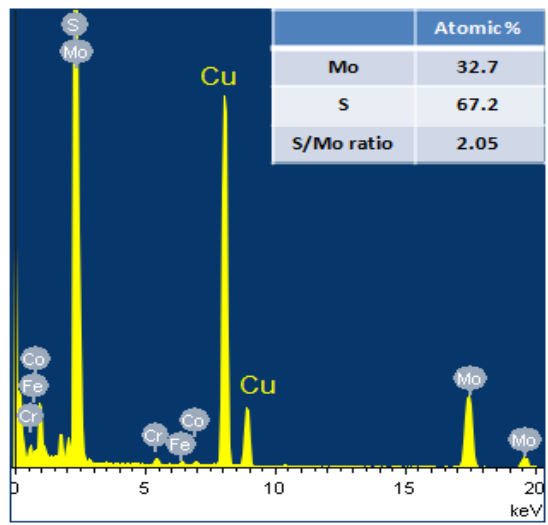

**Figure S5**. TEM-EDS spectrum of the $MoS_2$ film. Inset shows the stoichiometry of the $MoS_2$ film.

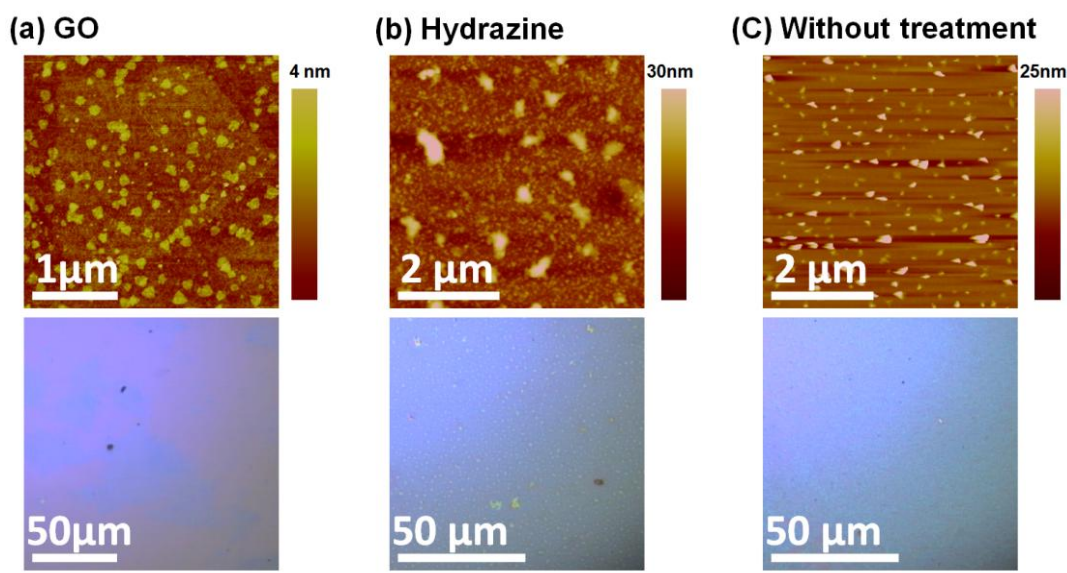

**Figure S6** Typical AFM and OM images of the MoS$_2$ sample deposited on the substrate subject to (a) the coating of graphene oxide (b) the immersion of hydrazine solution (c) no treatment. Only nanoparticles were formed on these substrates as evidenced by AFM measurements.

**Table S1.**

| Treatment | After growth |
|---|---|
| No treatment | MoS$_2$ Nanoparticles |
| Hydrazine | MoS$_2$ Nanoparticles |
| Graphene oxide (GO) | MoS$_2$ Nanoparticles |
| **Reduced graphene oxides (rGO)** | **Large-area layered MoS$_2$** |
| KCl | MoS$_2$ Nanoparticles |
| **perylene-3,4,9,10-tetracarboxylic acid tetrapotassium salt (PTAS)** | **Large area layered MoS$_2$** |
| **perylene-3,4,9,10-tetracarboxylic dianhydride (PTCDA)** | **Large area layered MoS$_2$** |

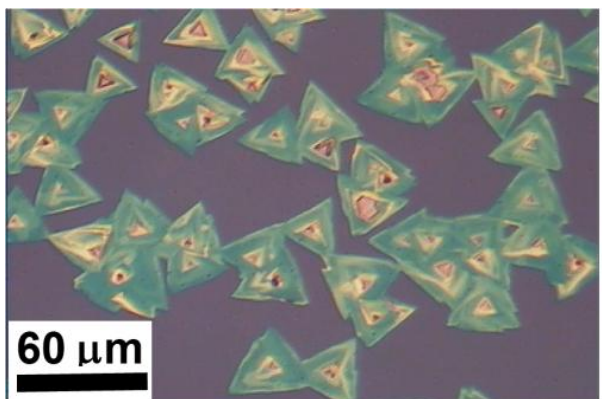 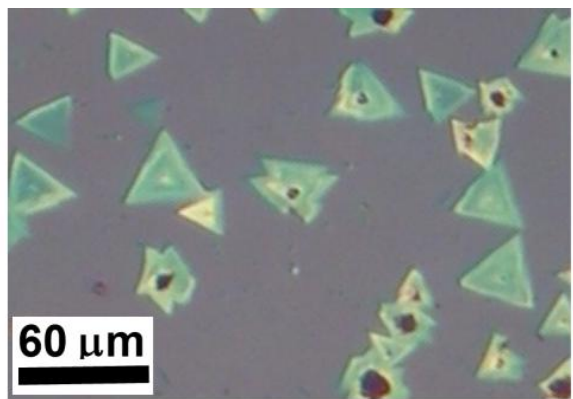

**Figure S7.** Optical micrographs showing the growth of WS$_2$ layers on the substrates pre-treated with PTAS and rGO.

9